# Barrier properties of k-mer packings

N. Lebovka[a,*], S. Khrapatiy[b], Vygornitskyi[a], N. Pivovarova[c]

[a]*Institute of Biocolloidal Chemistry named after F.D. Ovcharenko, NAS of Ukraine, 42, blvr. Vernadskogo, Kyiv 03142, Ukraine, lebovka@gmail.com (Lebovka N.)*
[b]*Taras Shevchenko National* University of Kyiv, *Institute of Biology, 2, prospekt Hlushkov, Kyiv 03022, Ukraine, khrapatiysv@ukr.net (Khrapatiy S.)*
[d]*National University of Life and Environment Sciences of Ukraine, 15 Geroiv Oborony, Kyiv 03041, Ukraine, natpiv@gmail.com (Pivovarova N.)*

## Abstract

This work discusses numerical studies of the barrier properties of *k*-mer packings by Monte Carlo method. The studied variants of regular and non-regular arrangements on a square lattice included models of random sequential adsorption (RSA) and random deposition (RD). The discrete problem of diffusion through the bonds of a square lattice was considered. The *k*-mers were perfectly oriented perpendicular to the diffusion direction and blocked certain fraction of bonds $f_b$ against diffusion. The barrier efficiency was estimated by calculation of the ratio $D/D_o$ where $D$ is diffusion coefficient in direction perpendicular to the orientation of *k*-mers and $D_o$ is the same value for diffusion on the square lattice without blocked bonds, i.e., at $f_b=0$. The value of $k$ varied from 1 to 512 and different lattice sizes up to $L$=8192 lattice units were used. For dense packings ($p$=1), the obtained $D/D_o$ versus $f_b$ dependences deviated from the theoretical prediction of effective medium (EM) theory and deviation was the most obvious for the regular quadratic arrangement. For loose RSA and RD packings, the percolation like-behavior of $D/D_o$ with threshold at $f_b=p_\infty$ was observed and the data evidenced that their barrier properties at large values of $k$ may be more effective than those of some dense packings. Such anomalous behavior can reflect the details of *k*-mer spatial organization (aggregation) and structure of pores in RD and RSA packings. The contradictions between simulation data and predictions of EM theory were also discussed.

## 1. Introduction

Barrier properties of films, filled by impermeable particles of anisotropic shape, continuously attract both fundamental and practical interests[1–5]. The possible applications include barrier film for gas tanks, coatings, films for packaging food, carbonated drinks and hazardous wastes. The most common fillers are the phyllosilicate particles (mica, montmorillonite, Laponite and vermiculite) with high aspect ratio *k* (length/thickness ratio). They can noticeably reduce film permeability even at small loadings (≤1%). The barrier effect may be explained by formation of more complicated tortuous paths for the diffusion molecules in the presence of a filler. The relative diffusion coefficient (ratio of diffusion coefficients of film in the presence, *D*, and in the absence, $D_o$, of filler may be expressed as[2,3]

$$D=D_o/\tau, \tag{1}$$

where $\tau$ is tortuosity defined as

$$\tau=L/L_o>1, \tag{2}$$

---

*Corresponding author. Fax: +380 444280378.
*E-mail address:* lebovka@gmail.com

where $L$ and $L_o$ are the diffusion distance for solute molecule in the presence and the absence of anisotropic filler.

Different permeability models[6–10] were proposed for estimation of tortuosity $\tau$. The simplest models consider arrangements of rectangular particles with orientation that is perpendicular to the diffusion direction. For this case, Nielsen used effective medium (EM) approximation and obtained[6]:

$$D/D_o = (1-p_v)/(1+kp_v/2) \qquad (3)$$

where $p_v$ is the packing density, or volume fraction, of barriers in the film.

In a general case, the $D/D_o$ ratio may depends also on many details of particle arrangements inside the film, e.g., the shape of pores, particle agglomerations, orientation and etc[2,3]. Numerical simulation of diffusion through films, containing anisotropic particles, allows taking of these details into account and checking of the consistence of geometry-based equations for diffusion coefficient. Up to now several works on numerical simulation[10–17] were done for this diffusion problem. However, the results were obtained for the limited number of geometrical configurations and at small scales and were not generally sufficient for making an adequate comparison between the theory and data of numerical simulations. The large-scale simulations[13] have shown that Nielsen model was not accurate enough and the following equation was obtained for the continuous placement of platelets in the space:

$$D/D_o = 1/(1+c_1 k p_v + c_2 (k p_v)^2), \qquad (4)$$

where $c_1$ and $c_2$ are the numerical coefficients. In two dimensions, $c_1=0.46\pm0.01$ and $c_2=0.165\pm0.01$. It was shown that this relation worked remarkably well over the entire range of inclusion concentrations, even close to the dense packing limit and the more remarkable is that the quadratic term dominated for large aspect ratio, $k\gg1$,.

The present work utilised the discrete model for calculation of diffusion coefficient. The barrier particles were represented by linear $k$-mers, i.e., the particles occupying $k$ near-neighbor sites on a square lattice. Note that the models of $k$-mers were frequently used to study of packing and jamming[18–21], percolation[22–24], phase transitions[25] and diffusion[26] phenomena. The remainder of the paper is organized as follows. In Section II, we describe our model and the details of simulation. The obtained results are discussed in Section III. We summarize the results and conclude our paper in Section IV.

## 2. Computations

The different variants of $k$-mer packings on a square lattice were used for simulation of the structure of films with barrier inclusions. It was supposed that an empty square lattice consisted of vertical and horizontal bonds. The $k$-mers served as the diffusion barriers. They were placed with their long axis parallel to the horizontal direction ($x$-axis), i.e., the problem of perfectly oriented $k$-mers was considered.

### 2.1. Models of packing of k-mers

The studied variants of regular and non-regular packings of $k$-mers are presented in Fig. 1. Here, the packings $P_0$ and $P_{k/2}$ represent the regular quadratic and staggered arrangements, respectively, and packing $P_r$ represents the random arrangement of $k$-mers in the neighbor layers. The blocked bonds on the square lattice are shown by crosses. Note that in $P_{k/2}$ packing, the neighbor layers are shifted on the $k/2$ site units. The packings $P_1$ and $P_2$ that are shifted on the 1 and 2 site units were also studied in this work. All packings $P_0$, $P_{k/2}$, $P_r$, $P_1$ and $P_2$ are dense, i.e., their packing density is $p=1$. The RSA packing (Fig.1) was obtained using the jamming state of the one-dimensional model of random sequential adsorption (RSA)[27] for each horizontal layer of $k$-mers. The RSA packing is loose, and packing density $p$ in the jamming state may be calculated according to Krapivsky *et al.*

$$p = k \int_0^\infty du \exp\left(-u - 2\sum_{i=1}^{k-1}(1-\exp(-iu))/i\right) \tag{5}$$

For $k \to \infty$, the packing density tends to Renyi's parking constant $p \to c_R \approx 0.7475979202$[27] (for numerical values at different k see[19]).

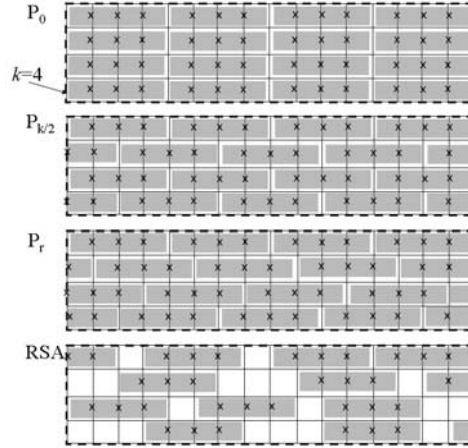

Fig. 1. Different models of packing: low ($P_0$) and high tortuous ($P_{k/2}$, $P_r$) dense packings and RSA loose packing of k-mers. The blocked bonds on the square lattice are shown by crosses.

As a variant of loose packing, the random deposition (RD) packing was also studied. Note that different variants of models for formation of random deposits from anisotropic particles were previously developed. E.g., 1+1 RD models were developed for simulation of the structure of loess deposits[28,29]. The 'Loughborough Loess' deposits, formed by anisotropic particles, exhibited the complex porous structure and had high porosity. Computer experiments with RD model for mixture of monomers (k=1) and dimers (k=2) have shown that deposit density $p$ was dependent on the fraction of dimers[30–32]. It was noticed that the pores were anisotropic and the connectivity was larger in the vertical direction. It was also demonstrated that the sample spanning pore had a fractal structure[32]. The relaxation process that allows the particles to rearrange into a more stable configurations (it mimics the process of compaction) has noticeable effect on the structure of such mixed deposits[33–36].

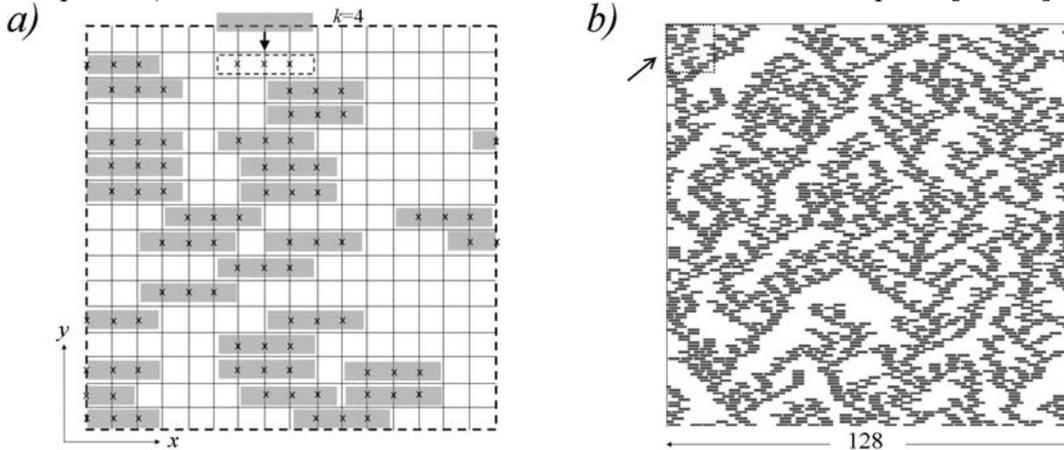

Fig. 2. Description of on-top site random-deposition (RD) model (a) and example of deposit (b) for k=4. The blocked bonds on the square lattice are shown by crosses. Here, a) is an enlarged portion of the pattern b) shown by an arrow.

In our case, the simple on-top site 1+1 RD model was used. The k-mers were deposited sequentially with their long axis parallel to the horizontal direction (x-axis) and the vertical direction (y-axis) coinciding with the direction of deposition (Fig. 2a).

The periodical boundary conditions were applied in the horizontal direction. For each next $k$-mer, the height of all columns, were it was dropped, were checked, and then it was added to the top of the highest column. The example of RD for $k=4$ is presented in Fig. 2b.

For the case of monomer, $k=1$, this RD model coincides with the simple random deposition model that gives the dense packing, $p=1$. The local density of sediment $p_h$ in the raw $y=h$ is calculated as a ratio of the filled sites in the raw and the size of the lattice is $L_x=L$. The values of $p_h$ were calculated for completely saturated layers. Preliminary experiments have shown that it can be obtained by choosing the sufficiently large size of the system in the vertical direction $L_y=h+L$.

### 2.2. Calculation of diffusion coefficient

The discrete problem of diffusion through the bonds of a square lattice was considered. The barrier efficiency was estimated by calculation of diffusion coefficient $D$ in the direction perpendicular to the orientation of $k$-mers (horizontal direction). The $k$-mers, placed on the square lattice, blocked the vertical bonds and acted as a barrier. In the absence of $k$-mers, the problem corresponded to the simple diffusion through the bonds of a square lattice. The method of random walks of the trial particle with initial position in the centre of the square lattice of size $LxL$ was used. The periodical boundary conditions were applied in the horizontal direction.

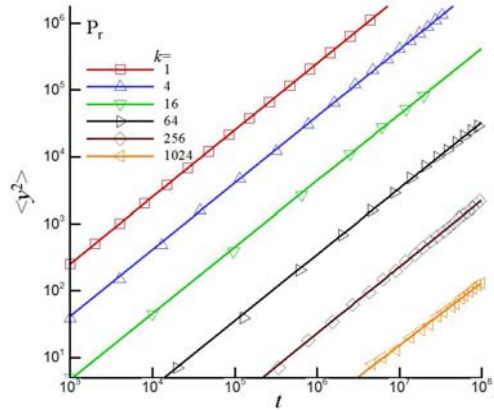

Fig. 3. Mean square displacement in the vertical direction $<y^2>$ versus diffusion time $t$ for $P_r$ packing at different values of $k$.

The examples of mean square displacement in the vertical direction $<y^2>$ versus the diffusion time $t$ (number of displacements) for diffusion on $P_r$ packing are presented in Fig. 3.

For all the studied packings of $k$-mers, the diffusion was normal and the value of $<y^2>$ was linearly growing with time $t$, i.e.,

$D \propto <y^2>/t$. (6)

Finally, the results were presented as the ratio $D/D_o$, where $D_o$ corresponds to diffusion on a square lattice without blocked bonds.

### 2.3 Other details

The correspondence between continuous and discrete variants of diffusion problems may be established as follow. The volume fraction of impermeable barrier, $p_v$, in continuous diffusion problems (see, eq. (3), (4)) is equivalent to the fraction of blocked vertical bonds, $f_b$, in the lattice diffusion problem, i.e. $p_v=f_b$. For the problem of $k$-mers under consideration, each $k$-mer blocks $k-1$ vertical bonds and the value of $f_b$ may be calculated as

$f_b = p(k-1)/k = p(1-1/k)$. (7)

EM equation (Eq.(3)) may be rewritten for the bond diffusion problem as

$$D/D_o = (1-p_b)/(1+kp_b/2) = (1-p(1-1/k))/(1+p(k-1)/2), \quad (8)$$

and in the limit of long k-mers ($k\to\infty$) it gives

$$D/D_o = 2/(k(k+1)) \xrightarrow[k\to\infty]{} 1/c_{EM}k^2 \quad (9)$$

for dense packings (here, $c_{EM}$=0.5 and $p$=1), and

$$D/D_o \xrightarrow[k\to\infty]{} 2(1-p)/pk \quad (10)$$

for loose packings ($p$<1).

Thus, EM theory predicts the following power law dependence:

$$D/D_o = 1/ck^\gamma \quad (11)$$

with crossover between $\gamma \approx 2$ for dense packings (p=1) and $\gamma \approx 1$ for loose packings ($p$<1). Here, $c$ is a coefficient.

In all computer calculations, the maximum size of the square lattice $L$ was of 8192 lattice units. The data were averaged using 1000 independent runs.

## 3. Results and discussion

*3.1. Packing density versus k-mer length for RD model*

The study of local packing density $p_h$ versus the vertical position $h$ and size of the lattice $L=L_x$ was done as the first step (Fig.4). For the given $k$ (=2-256), the value of $p_h$ decreased as $h$ decreased and got saturated at large $h$.

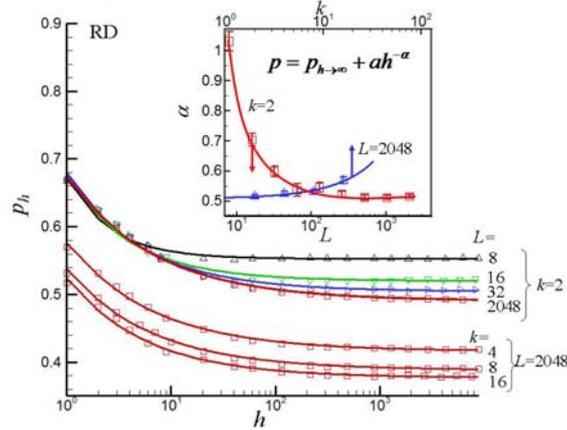

Fig. 4. Examples of local packing density $p_h$ versus vertical position $h$ for different values of $k$ and $L$. Insert shows examples of dependences of the power exponent in Eq. (12) $\alpha$ versus $L$ for $k$=2 and $\alpha$ versus $k$ for $L$=2048.

Analysis has shown that $p_h(h)$ dependences may be good fitted using the following scaling relation

$$p_h = p_{h\infty} + a/h^\alpha \quad (12)$$

where $p_{h\infty}=p_h(h\to\infty)$, $a$ and $\alpha$ are the scaling parameters.

The scaling parameters $a$ and $\alpha$ were dependent on $k$ and $L$. Insert on Fig. 4 shows examples of dependences of the power exponent $\alpha$ versus $L$ for $k$=2 and $\alpha$ versus $k$ for $L$=2048. Note that small value of $\alpha$ corresponds to slow regime of $p_h$ approach to the saturation value $p_{h\infty}$. The value of $\alpha$ decreased as $L$ increased, which corresponds to slower regime for larger values of $L$. From the other side, faster regime was observed for larger values of $k$.

In fact, the saturation value of local packing density $p_{h\infty}$ corresponds to the mean packing density of the deposit $p$ at large values of $h$, i.e., $p=p_{h\infty}$. Note that portions of RSA packings with $p \approx p_{h\infty}$ at large values of $h$ were used in the following calculations of $D/D_o$ values.

Finally, *p* versus *L* dependency was used for determination of the thermodynamic values of $p_\infty(k)$ in the limit of $L\to\infty$. Analysis has shown that computational data can be good fitted using the following scaling relation

$p(k)=p_\infty(k)+bk/L$        (13)

where *b* is the scaling parameters.

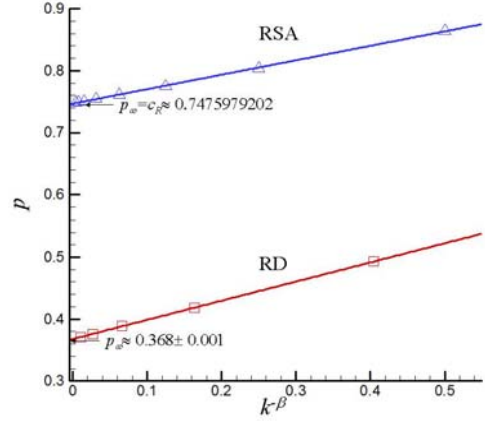

Fig. 5. Packing density *p* versus values of $1/k^\beta$ for packings corresponding to the RD ($\beta=1.306\pm0.024$) and RSA($\beta\approx1.0$) models. The limiting values of packing fraction at $k\to\infty$ are $p_\infty=0.368\pm0.01$ and $p_\infty=c_R\approx0.7475979202$ (Renyi's parking constant)[27] for RD and RSA models, respectively.

Figure 5 shows the thermodynamic values of mean packing density of deposit *p* versus values of $1/k^\beta$ for the packing, corresponding to RD model. The value of $\beta$, adopted by rms fitting, was $1.306\pm0.024$, and the limiting value of packing fraction at $k\to\infty$ was $p_\infty(RD)=0.368\pm0.01$., The analogical dependence with $\beta\approx1$ and $p_\infty(RSA)=c_R\approx0.7475979202$ (Renyi's parking constant) for packing, corresponding to the RD model, is also presented here for completeness. The obtained data evidence the simple power law relation between *p* and *k*:

$p=p_\infty+d/k^\beta$        (14)

where $d=0.309\pm0.002$ and $d=0.228\pm0.002$ for RD and RSA models, respectively.

*3.1. Behaviour of diffusion coefficients for different packings*

Figure 6 shows the relative diffusion coefficient $D/D_o$ versus the fraction of blocked vertical bonds $f_b$. The data were obtained for different packings. For all dense packings ($p=1$), the obtained $D/D_o$ versus $f_b$ dependences deviated from theoretical prediction of EM theory (dashed line in Fig. 6, Eq. (8)).

The most obvious was deviation observed at regular quadratic arrangement ($P_0$). For this arrangement, the permeability was maximal at the given $f_b$ and near linear correspondence was observed between $D/D_o$ and $f_b$. For random arrangement of *k*-mers in the neighbor layers ($P_r$), the correspondence between computer "experiments" and EM theory was rather close. On the contrary, the deviation between "experiments" and EM theory was larger for loose packings (RSA and RD). For these packings, the value of $D/D_o$ was very small above the threshold fraction of blocked vertical bonds $f_b=p_\infty$ and became noticeable below it. Hence, this behaviour was rather similar to the percolation behavior with a threshold at $f_b=p_\infty$.

Fig. 7 shows dependences of $D/D_o$ on *k* for different packings. For dense packings ($p=1$), the power law dependence, expressed by Eq. (11), was observed in the limit of $k\to\infty$ with $\gamma\approx1$ for $P_0$, $P_1$, $P_2$ packings and $\gamma\approx2$ for $P_{k/2}$, $P_r$ packings. The inverse quadratic dependencies with $\gamma\approx2$ for high tortuous packings $P_{k/2}, P_r$ corresponded to the prediction of EM theory in the limit of $k\to\infty$, Eq.(9). From the other side, the regimes with $\gamma\approx1$ were observed for packings $P_0$, $P_1$ and $P_2$. It was in evident contradiction with EM theory that predict $\gamma \approx 1$ for all dense packings ($p=1$)

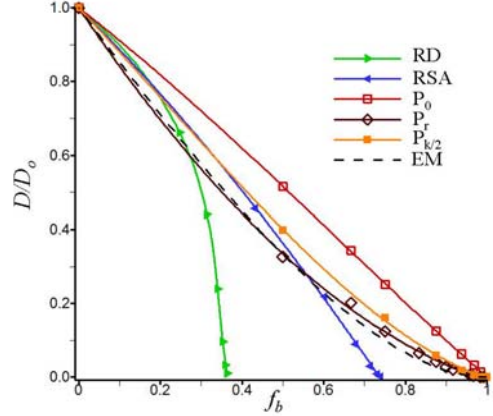

Fig. 6. Relative diffusion coefficient $D/D_o$ versus fraction of blocked vertical bonds $f_b$. The data were obtained for different packings, and dashed line corresponds to the EM theory, Eq. (8).

Moreover, it was observed that $D/D_o$ values of loose RD and RSA packings at small $k$ may exceed the same values for dense $P_0$, $P_1$ and $P_2$ packings. However, the opposite behaviour was observed above certain threshold value of $k_c$: for $k \geq k_c$ the barrier efficiency of the loose packing was better than those of low tortuous packings $P_0$, $P_1$ and $P_2$. E.g., the threshold values were $k_c \approx 32(P_0)$, $k_c \approx 64(P_1)$ and $k_c \approx 128(P_2)$ for RD deposit and they were $k_c \approx 2(P_0)$, $k_c \approx 16(P_1)$ and $k_c \approx 32(P_2)$ for RSA packing.

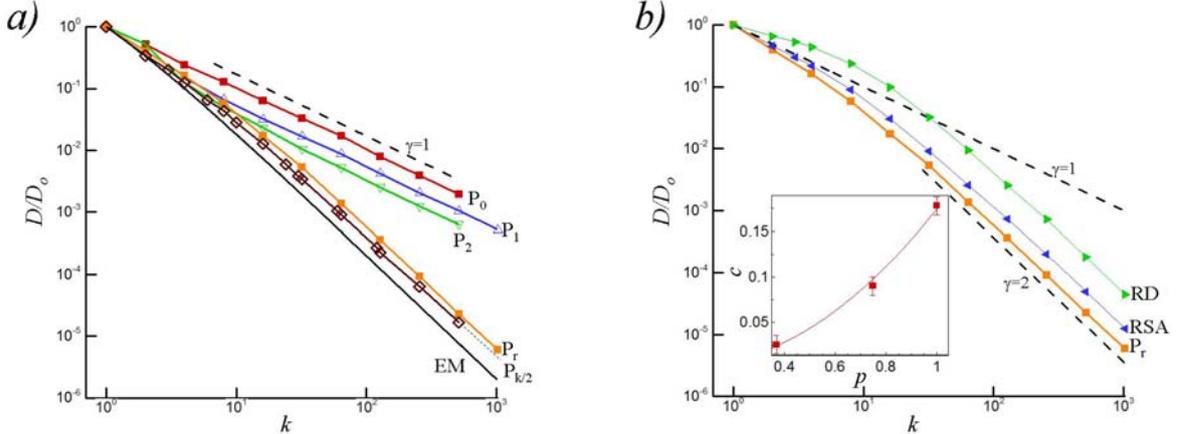

Fig. 7. Relative diffusion coefficient $D/D_o$ versus length of $k$-mer. Comparison of different dense packings (a) and of loose and dense packings (b). Solid line EM corresponds to the effective medium theory, Eq.(9). Dashed lines correspond to $\gamma \approx 1$ and $\gamma \approx 2$ in Eq. (14). Insert in Fig. 6b shows $c(p)$ dependence of the coefficient in Eq. (14), obtained from data for RD ($p \approx 0.368$), RSA ($p \approx 0.748$), and $P_r$ ($p=1$) packings.

Such behaviour is surprising and can reflect the details of $k$-mer spatial organization (aggregation) and structure of pores in loose RD and RSA packings at large values of $k$. The calculated value of $D/D_o$ represents the weighted mean value averaged over different places with high (pores) and low (aggregates of $k$-mers) permeability. The observed anomalies in barrier properties of RD and RSA packings at large $k$ may reflect the high relative contribution of places with low permeability (i.e., places represented by $k$-mer aggregates) to the value of $D/D_o$. For RD and RSA packings, the value of $D/D_o$ followed the power law regime with $\gamma \approx 2$ in the limit of $k \to \infty$ (Eq. 11) and $c$ parameter was increasing as packing density $p$ increased. Insert in Fig. 6b presents $c$ versus $p$ dependence, obtained from data for RD ($p\approx0.368$), RSA ($p\approx0.748$), and $P_r$ ($p=1$) packings. Analysis has shown that $c(p)$ dependence may be well approximated by simple parabolic law:

$$c = c_2 p^2 \qquad (15)$$

where $c_2=0.179\pm0.006$.

Finally, Eq.(14) and Eq. (15) may be combined as
$$D/D_o=1/c_2(pk)^2. \qquad (16)$$

This equation for the discrete bond diffusion problem is of the same form as was previously proposed for the continuous model of aligned platelets in the limit of $k\to\infty$. Moreover, the obtained numerical values of eq.16 coefficients for discrete and continuous models, $c_2=0.179\pm0.006$ and $c_2=0.165\pm0.01$[13], respectively, were in reasonable correspondence. Note that diffusion behaviour observed for discrete lattice models was in contradiction with EM theory, which predicts in the limit of long $k$-mers ($k\to\infty$) crossover between $\gamma\approx2$ for all dense packings and $\gamma\approx1$ for loose packings ($p<1$).

It is interesting to test in future investigations the universality of Eq. (16) for different other variants of $k$-mer parkings in the space accounting for their stacking and orientation order[19,24].

## 4. Conclusions

In this paper, the barrier properties of two dimensional films, filled by perfectly oriented anisotropic particles ($k$-mers), were investigated by computer simulations. The $k$-mers were placed on the square lattice using different variants of regular and non-regular arrangements for dense packings and models of random sequential adsorption (RSA) and random deposition (RD) for loose packings. The discrete problem of diffusion through the bonds of the square lattice was considered. The barrier efficiency was estimated by calculation of the $D/D_o$ ratio, where $D$ is the diffusion coefficient in direction perpendicular to the orientation of $k$-mers and $D_o$ is the same coefficient for diffusion on a square lattice without blocked bonds. The value of k varied from 1 to 512 and different lattice sizes up to $L=8192$ lattice units were used. For RD packings, more detailed studies of the scaling properties of packing density $p$ were done and power law relation $p= p_\infty+d/k^\beta$ with $p_\infty=0.368\pm0.01$, $d=0.309\pm0.002$, $\beta=1.306\pm0.024$ was observed. The $D/D_o$ versus the fraction of blocked vertical bonds $f_b$ and the value of $k$ were compared for different packings with prediction of effective medium (EM) theory. The obtained $D/D_o$ versus $f_b$ dependences deviated from EM theoretical prediction for all dense packings ($p=1$) and deviation was the most obvious for the regular quadratic arrangement. For loose RSA and RD packings, the percolation like-behavior of $D/D_o$ with the threshold at $f_b=p_\infty$ was observed. In the limit of $k\to\infty$ the power law dependences $D/D_o\propto k^{-\gamma}$ for dense packing (with $\gamma\approx1$ and $\gamma\approx2$) and loose packings ($\gamma\approx2$) were observed. It evidences that barrier properties of loose RSA and RD packings can be higher than those of some dense packings at large values of $k$. Such anomalous behavior may reflect the details of $k$-mer spatial organisation (aggregation) and structure of pores in RD and RSA packings. Such behaviour is in contraction with the EM theory that predicts crossover between $\gamma\approx2$ for dense packings ($p=1$) and $\gamma\approx1$ for loose packings ($p<1$).

## Acknowledgment

The authors would like to acknowledge the partial financial support within the project N60/13-H of program "Fundamental problem of nanostructured systems, nanomaterials, nanotechnologies", Ukraine.